\documentclass[10pt,letterpaper,twocolumn]{article}

\usepackage[letterpaper,
  top=0.98in,
  bottom=0.98in,
  left=0.98in,
  right=0.98in,
  headheight=22pt,
  headsep=8pt,
  footskip=24pt]{geometry}

\usepackage[T1]{fontenc}
\usepackage{newtxtext,newtxmath} 
\usepackage{microtype}
\usepackage{graphicx}
\usepackage{booktabs}
\usepackage{array}
\usepackage{amsmath}
\usepackage{enumitem}
\usepackage{titlesec}
\usepackage{fancyhdr}
\usepackage{lastpage}
\usepackage{caption}
\usepackage{url}
\usepackage[numbers,sort&compress]{natbib}
\usepackage[hidelinks]{hyperref}

\newcommand{\IACPaperCode}{IAC-26-A4.1.6}
\newcommand{\IACTitle}{The Robots are Coming: Artificial Intelligence Searching for Extraterrestrial Intelligence}
\newcommand{\IACConference}{Presented at the 77\textsuperscript{th} International Astronautical Congress (IAC 2026), Antalya, T\"urkiye, 5--9 Oct 2026.}
\newcommand{\IACCopyright}{}

\titleformat{\section}
  {\normalfont\bfseries\fontsize{10}{11}\selectfont}
  {\thesection.}{0.45em}{}
\titlespacing*{\section}{0pt}{9pt}{1pt}

\titleformat{\subsection}
  {\normalfont\itshape\fontsize{10}{11}\selectfont}
  {\thesubsection}{0.45em}{}
\titlespacing*{\subsection}{0pt}{7pt}{1pt}

\titleformat{\subsubsection}
  {\normalfont\itshape\fontsize{10}{11}\selectfont}
  {\thesubsubsection}{0.45em}{}
\titlespacing*{\subsubsection}{0pt}{6pt}{1pt}

\newcommand{\iacunnumbered}[1]{%
  \par\vspace{7pt}%
  \noindent\textbf{#1}\par\vspace{1pt}}

\setlist{nosep,leftmargin=*}

\begin{document}

\twocolumn[{%
\begin{minipage}{\textwidth}
\begin{center}
  \fontsize{11}{12}\selectfont
  \IACPaperCode\

  \vspace{8pt}

  {\bfseries\fontsize{12}{14}\selectfont \IACTitle}

  \vspace{8pt}

  {\bfseries\fontsize{11}{13}\selectfont
  Steve Croft\textsuperscript{a,b,c}}
\end{center}

\vspace{5pt}

\noindent\textsuperscript{a}\textit{Breakthrough Listen, University of Oxford, Denys Wilkinson Building, Keble Road, Oxford OX1 3RH, UK}\par
\vspace{-1pt}
\noindent\textsuperscript{b}\textit{Breakthrough Listen, University of California, Berkeley, 501 Campbell Hall \#3411, Berkeley, CA 94720, USA}\par
\vspace{-1pt}
\noindent\textsuperscript{c}\textit{SETI Institute, 339 Bernardo Ave, Suite 200, Mountain View, CA 94043, USA}

\vspace{8pt}

\vspace{11pt}

\begin{center}
\textbf{Abstract}
\end{center}
\vspace{-4pt}

\noindent
Breakthrough Listen, humanity's most ambitious search for technosignatures, or evidence of extraterrestrial technology, is carrying out surveys across the sky and across the electromagnetic spectrum, in images, spectra, time series, catalogs, and alerts. Classical techniques still play a major role in the search, but the recent rapid growth in the capabilities of artificial intelligence is enabling new and more powerful algorithms and approaches to be deployed across enormous datasets. Surveys designed to look for technosignatures are still a major focus for the Breakthrough Listen program, but excitingly, the new strategies also enable comprehensive mining of archives and of general-purpose real-time data streams.

We outline how artificial intelligence and machine learning are being used to scale up the search, including with the newly-commissioned Vera C. Rubin Observatory, in addition to radio telescopes around the planet. The chances of a detection have never been better.

\vspace{2pt}

\vspace{8pt}
\end{minipage}
}]


\iacunnumbered{Acronyms/Abbreviations}
\noindent\begin{tabular}{@{}p{0.28\linewidth}@{}p{0.72\linewidth}@{}}
AHA & Anomaly Hunter for Alerts \\
AI & artificial intelligence \\
APF & Automated Planet Finder \\
ATA & Allen Telescope Array \\
BLADE & Breakthrough Listen Accelerated Digital Signal Processor (DSP) Engine \\
BLUSE & Breakthrough Listen User-Supplied Equipment \\
ETI & extraterrestrial intelligence \\
FRB & fast radio burst \\
FRBNN & fast radio burst neural network \\
GBT & Green Bank Telescope \\
GLOBULAR & Grouping Low-frequency Observations By Unsupervised Learning After Reduction \\
GPS & global positioning system \\
GPU & graphics processing unit \\
HDBSCAN & hierarchical density-based spatial clustering of applications with noise \\
ISO & interstellar object \\
LLM & large language model \\
LSST & Legacy Survey of Space and Time \\
LSTM & Long Short-Term Memory \\
ML & machine learning \\
RFI & radio frequency interference \\
SDSS & Sloan Digital Sky Survey \\
SETI & search for extraterrestrial intelligence \\
TESS & Transiting Exoplanet Survey Satellite \\
\end{tabular}

\noindent\begin{tabular}{@{}p{0.28\linewidth}@{}p{0.72\linewidth}@{}}
UMAP & Uniform Manifold Approximation and Projection \\
ZTF & Zwicky Transient Facility \\
\end{tabular}

\section{Introduction}

Technosignatures, or evidence of extraterrestrial technology as a proxy for extraterrestrial intelligence, could take many forms.
Early searches focused on narrow-band radio transmissions \cite{drake1961}, but technosignatures could also encompass broad-band radio signals;
laser lines in stellar spectra; dimming of stars, or excess infrared emission, due to megastructures; artificial objects in the Solar System; or technologies that no Earthling has yet imagined.

Searches for known classes of signals can use filtering techniques that are matched to expected signal morphologies. For example, Listen's workhorse Doppler search pipeline, turboSETI \cite{turboseti}, is optimized for the detection of drifting narrow-band signals in radio spectrograms, although it detects other signal morphologies (including signals such as GPS) with lower efficiency. But ``unknown unknowns'' are natural targets for more generalized anomaly detection algorithms, which is where machine learning (ML) and artificial intelligence (AI) come to the forefront.

The prevalence, luminosity distribution, intermittency, beaming, frequency, and bandwidth of technosignatures are unknown, making SETI a ``needle in a haystack'' problem \cite{wright2018}. To find the needle, we may need to sift through a larger quantity of hay. But it is also possible that needles are quite common, and that we just need a better method of finding them. To use another analogy from Jill Tarter, sampling one glass of seawater and concluding that there are no fish in the ocean would be foolish. However, one glass of water contains plenty of life -- the searcher just needs to put some of the water under a microscope to see it. The new AI-driven methods may provide the tools we need to find technosignatures in existing datasets, at the same time new surveys are expanding the search volume. 

These AI and ML methods take several approaches. Trained classifiers can find signal classes that we already know how to describe, or can reject interference or instrumental artifacts. Anomaly detectors can tag observations that are unlike the background population without being told what to look for. And, more recently, large language models (LLMs) are showing promise in triaging candidates generated from classical search pipelines, and may be able to assist with human review in other areas as well. These ML approaches do different jobs, cost different amounts of money to run, and need to be applied and validated in different ways.

Breakthrough Listen \cite{worden2017}, the largest SETI program to date, uses telescopes and archives around the world (and in space) and across the electromagnetic spectrum, applying a wide variety of techniques to search for a wide range of possible technosignatures. In this paper I review the development of AI and ML approaches within the Listen collaboration. I also briefly cite some relevant work from outside of Listen, but this paper is not intended as a comprehensive review of the anomaly detection or SETI literature.

Section~2 outlines classical approaches that have been applied to Listen data (both radio and optical) to date.
Section~3 covers ML analyses that have been performed on our radio data.
Section~4 covers anomaly detection and interference rejection,
Section~5 describes real-time processing at the telescope,
and Section~6 some early work with LLMs.
Section~7 outlines ML analyses and anomaly detection with lightcurves, alert streams, and the next generation of optical surveys.
In Section~8 I draw some conclusions.

\section{Classical Analyses}

\subsection{Radio}

At Listen's radio facilities we record raw voltages, which pass through a GPU-accelerated Fourier Transform code to produce 
dynamic spectra (also referred to as spectrograms or waterfall plots): arrays of integrated power as a function of frequency and time. 
The standard pipeline at the single dish facilities, including the Green Bank Telescope (GBT) and Parkes, produces a high-spectral-resolution
spectrogram for narrow-band searches, a high-time-resolution one for pulse searches, and an intermediate one for spectral lines \cite{lebofsky2019}.

The classical radio search looks for narrow-band signals that drift in
frequency. Extremely narrow spectral features are not expected from known natural astrophysical emission processes, so an extraterrestrial entity might choose to concentrate energy in a narrow spectral window 
to stand out from the astrophysical background, and to improve contrast with background noise. We expect an extraterrestrial signal to drift because a transmitter on a rotating or orbiting exoplanet, observed from a rotating and orbiting Earth, will exhibit a Doppler shift that changes with time \cite{li2023}.

More of a problem than astrophysical contaminants is radio frequency interference (RFI) from human technology (including transmitters local to the telescope, and satellites). Our single-pixel receivers at single-dish facilities mitigate this by 
using an ``ON/OFF'' cadence: point at the target
for five minutes, point at a comparison position for five minutes, and repeat twice \cite{enriquez2017}. A signal that shows up only in
the ON pointings is behaving like something localized on the sky, whereas one that shows up in both is not.

\begin{figure}[htbp]
  \centering
    \includegraphics[width=\columnwidth,angle=180]{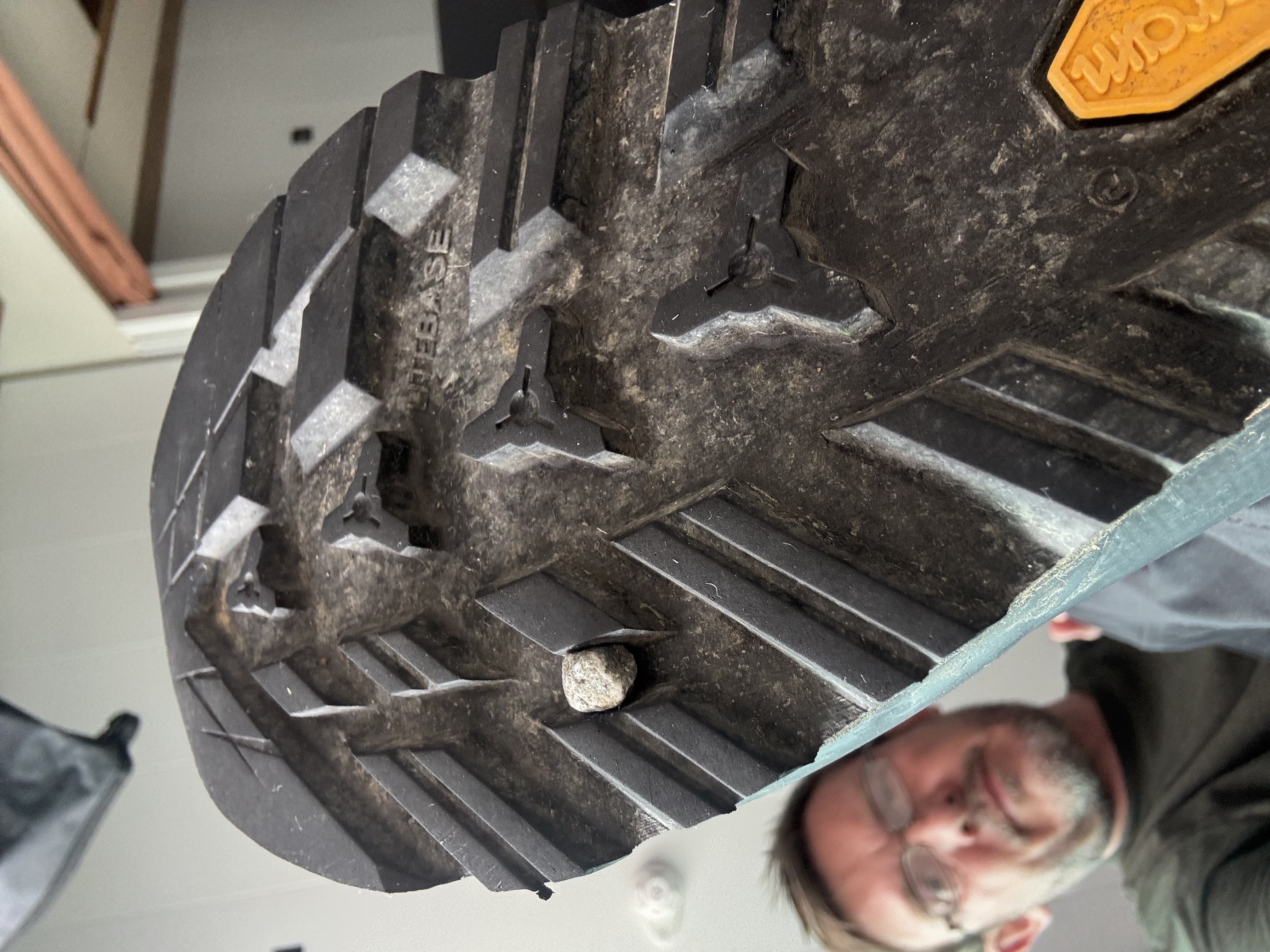}
  \caption{A piece of gravel stuck in the author's shoe is only interesting, compared to the millions of other pieces of gravel on a pathway, insofar as its morphology matches the morphology of the shoe's tread.}
  \label{fig:shoe}
\end{figure}

Passing the ON/OFF and Doppler filters earns a signal a closer look. However, a signal that passes both tests is still unlikely to be a bona fide technosignature. The matched filters cut down the volume of the haystack (possibly discarding some needles in the process) but what remains is still most likely hay (i.e. RFI). Given enough RFI signals (hundreds of millions in our largest searches) some will pass through. A stone stuck in your shoe is no more intrinsically interesting than the millions of other stones on a gravel driveway; it just happened to fit in the gaps in the tread (Fig.~\ref{fig:shoe}).

Large-N arrays offer Listen the opportunity to run commensally alongside other observations. 
At MeerKAT, the Breakthrough
Listen User-Supplied Equipment (BLUSE) backend processes the full available bandwidth from all antennas, forming coherent beams towards
whatever interesting objects happen to be in the field \cite{czech2026}. This provides a rapid and cost-effective way to scale the
search from thousands of objects at single-dish facilities to millions of objects. When the de-Doppler algorithm finds a signal in one 
of the beams, it triggers a dump of raw voltage data for later analysis. The analysis can be classical, as for the single-dish facilities, but the 
enormous number of targets requires triaging large numbers of signals. As data volumes grow from both single-dish and array facilities, effective anomaly
detection approaches become increasingly important.

\subsection{Optical\label{sec:optical}}

Listen's optical program to date has concentrated on laser lines which would manifest as emission features in stellar spectra too narrow to be astrophysical spectral lines. Lipman et al.\ carried out an early targeted search of this kind with the Automated Planet Finder (APF), pointing it at Boyajian's Star and finding no laser emission \cite{lipman2019}. Zuckerman et al.\ then generalized the approach into a pipeline that matches each APF spectrum against a library of empirical templates and searches the residual for emission features \cite{zuckerman2023}. 

Interest in optical lightcurves as a technosignature search space owes a good deal to Boyajian's Star, which was flagged by citizen scientists who noticed aperiodic dips of up to 20\% in its Kepler lightcurve that matched no known class of variable \cite{boyajian2016}.
A swarm of artificial objects was proposed as one explanation for the Boyajian's Star dips \cite{wright2016b}, but the dips turned out to be chromatic, ruling out an opaque megastructure \cite{boyajian2018}. Looking for dips in lightcurves is nevertheless a valid method of searching for technosignatures, and in order to scale this method to the next generation of surveys, ML may need to replace visual inspection by humans.

Systematic searches have identified additional main-sequence dippers. Tzanidakis et al.\ scored lightcurves for 63 million FGK stars in Gaia and ZTF and found 81 new dipper candidates, some of which look like versions of Boyajian's Star \cite{tzanidakis2025}. The most plausible mechanisms are circumstellar clumps and occulting companions with disks. This is the optical haystack: the vast majority of dips are likely to be caused by instrumental artifacts, by dust, or by companion objects, but a technosignature search needs to filter through these contaminants to find signals that defy astrophysical explanation.

Megastructures need not transit to be detectable. A civilization intercepting a large fraction of its star's output should re-radiate the energy in the infrared, so excess mid-infrared emission is a technosignature in its own right \cite{wright2014a,wright2014b}. Recent searches have combined Gaia, 2MASS, and WISE to look for stars with more mid-infrared flux than their photospheres can account for \cite{suazo2024,contardo2024}. The handful of candidates that survive are being scrutinized for contamination by background sources and circumstellar dust. Just as with radio searches, however, passing a filter is where the work starts, not where it ends.

A complementary approach constrains \emph{when} to look rather than what to look for. The SETI Ellipsoid identifies stars from which a transmitter, having seen a conspicuous event such as SN\,1987A, could have sent a signal timed to arrive at Earth now \cite{davenport2022}. Listen and collaborators have applied this framework to Gaia and TESS, examining lightcurves for changes in variability coincident with the time that each star intersects the expanding ellipsoid \cite{nilipour2023,cabrales2024}. No anomalies were found, but the method provides a physically-motivated prioritization of targets that can feed into an anomaly search.

Finally, the shape of a transit encodes the shape of what is transiting. Sandford and Kipping showed that a lightcurve can be inverted into an approximate silhouette of the occulter \cite{sandford2019}. Stacey et al.\ quantified how non-circular an occulter must be before TESS could tell it from a planet \cite{stacey2020}. The inversion is challenging, which is exactly the kind of problem that neural networks trained on large simulated libraries are suited to (Section~\ref{sec:optical_ml}).

\section{Machine Learning in Listen Radio Searches}

\subsection{Narrow-band}

The earliest ML work in radio technosignature searches started from a simple observation: a
spectrogram is an image, and by 2018 image classifiers were very good.
Harp et al.\ showed that this approach works for signal classification
\cite{harp2019}. Brzycki et al.\ developed this further, training
convolutional networks to localize faint narrow-band signals in
synthetic spectrograms containing strong interference, the regime where a
matched-filter search struggles \cite{brzycki2020}.

In 2021 Listen put a large set of GBT cadence snippets (small sections of frequency across all six scans in a cadence) on the data analysis competition website Kaggle \cite{kaggle2021}.
We injected simulated signals into a fraction of them, and invited anyone
to build a detector. More than 750 teams did so. The
winning entries were ensembles of image-classification networks with
aggressive data augmentation, which was informative but not
surprising. What was more instructive was the way the competition went
wrong: a leak in the original test set forced a reset partway through,
and after the reset, the winning solution exploited some subtleties in the method used for signal injection. Building a synthetic signal population that
transfers cleanly to real survey data turned out to be the hard part; with the exception of spacecraft downlinks such as Voyager, we have few 
true positive signals to work with.

Setigen exists to address this problem: it synthesizes radio
signals and injects them into real Listen data products
\cite{brzycki2022}. It can be used to test precision and recall of classical searches, and also to train ML models by labeling 
data with and without injected signals. Setigen is primarily designed for narrow-band radio technosignatures, however, and does not span the full range of possible signal morphologies.

Ma et al.\ \cite{ma2023} used setigen to inject signals into GBT data, training a variational autoencoder whose output fed into a random forest classifier, to 
distinguish between cadence snippets with signals only in ON observations (i.e. good candidates) and those with no clear distinction between ON and OFF scans (i.e. likely RFI). They used their trained algorithm to process 120\,TB of GBT observations of 820 nearby stars. The majority of the
data had already been searched classically using turboSETI \cite{enriquez2017,price2020} but Ma et al.\ turned up new candidates, including eight in cadences that had previously been analyzed. The candidates were not seen in reobservations and so could not be confirmed as technosignatures.

It is worth noting that classical approaches continue to develop. Painter et al.\ applied a new kurtosis-based algorithm to observations of 2841 targets from the GBT archive, recovering five of the candidates discovered by Ma et al. as well as some new ones
\cite{painter2024, painter_erratum}.

In addition, interstellar
scintillation \cite{brzycki2023} can provide a third filter (in addition to Doppler and sky localization) to reduce the haystack of RFI,  enabling classical approaches to be extended to larger datasets while keeping the number of false positives tractable.

Any ML method has to beat, or at least complement,
classical methods. Beating them means doing better on some
concrete measure: precision and recall of signal recovery, variety of signal morphologies detected, and / or hours of
human inspection saved.

Although the new analyses found no confirmed technosignatures, they did demonstrate that different methods 
generate different candidate lists when run on the same dataset. It is conceivable that a bona fide 
technosignature is present in the Listen data archive and that we just need to develop the right algorithm to find it. It is also worth noting
that reanalysis can, by definition, only be performed on data that has been archived. Streaming analysis tools (Section~\ref{sec:realtime}) must get
the prefiltering right in real time.

\subsection{Broadband Pulsed Beacons}

Instead of narrow-band sinusoids, ETI might choose to transmit 
broadband pulses. These are also energetically
cheap per unit of detectability, and could be made to look unambiguously artificial, for instance by applying a negative
apparent dispersion measure. Gajjar et al. \cite{gajjar2022} searched GBT
observations of 1,883 stars at 4--8\,GHz for 
artificially dispersed pulses, using a convolutional network as the
classifier. Although they found no technosignatures, they broadened the range of signal morphologies Listen looks for,
making for a more effective search of the data haystack.

\subsection{Astrophysical Discoveries in Listen Observations}

Although Listen has found no technosignatures to date, there are astrophysical signals of interest present in our data.
Zhang et al.\ \cite{zhang2018} trained a
convolutional network to look for pulses in GBT data of the repeating fast radio burst FRB\,121102 and found 72 pulses that had been missed in previous searches. Once again, no new observations were necessary -- just new methods to analyze data 
already in the archive.

\begin{figure}[htbp]
  \centering
       \includegraphics[width=\columnwidth]{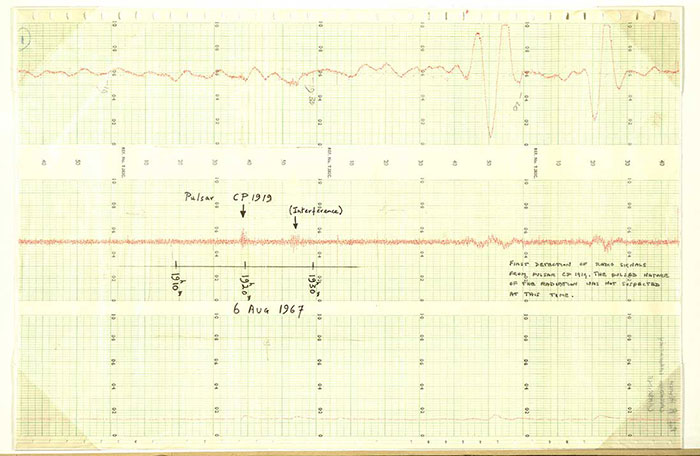}
\caption{``LGM1'', the first pulsar discovered, subsequently designated PSR\,B1919+21 (Churchill Archives Centre, Papers of Antony Hewish, HWSH Acc 355).}
  \label{fig:lgm}
\end{figure}

As previously noted, anomaly detection algorithms have been run on optical datasets and demonstrated the ability to find outlier objects that, while not necessarily technosignatures, are nevertheless interesting (see Section~\ref{sec:lightcurves}). The discovery of pulsars \cite{hewish1968} provides an excellent lesson; although they did not turn out to be signals from ``little green men'' (Fig.~\ref{fig:lgm}), their detection launched several new areas of modern astrophysics.

\section{Anomaly Detection and Interference Rejection}

\subsection{Grouping Signals After Detection}

One place to look for anomalies is downstream of the classical search,
in the population of hits it produces. Jacobson-Bell et al.\ took this
route with GLOBULAR, which applies HDBSCAN, a hierarchical
density-based clustering algorithm, to the hit population, groups
morphologically-similar interference together, and reduces the volume of hits that are fed into the ON/OFF event detection
\cite{jacobsonbell2025}. 

Using the same hits found in Choza et al.'s turboSETI search of 97 nearby
galaxies \cite{choza2024}, GLOBULAR cut false-positive hits by 93.1\%
and false-positive events by 99.3\% \cite{jacobsonbell2025}, while maintaining recovery of injected signals only slightly lower than the turboSETI-only method.
As the number of targets observed continues to grow, the number of candidates generated is becoming intractable for human inspection, making
additional filtering methods such as GLOBULAR essential.

\begin{figure*}[htbp]
  \centering
       \includegraphics[width=\textwidth]{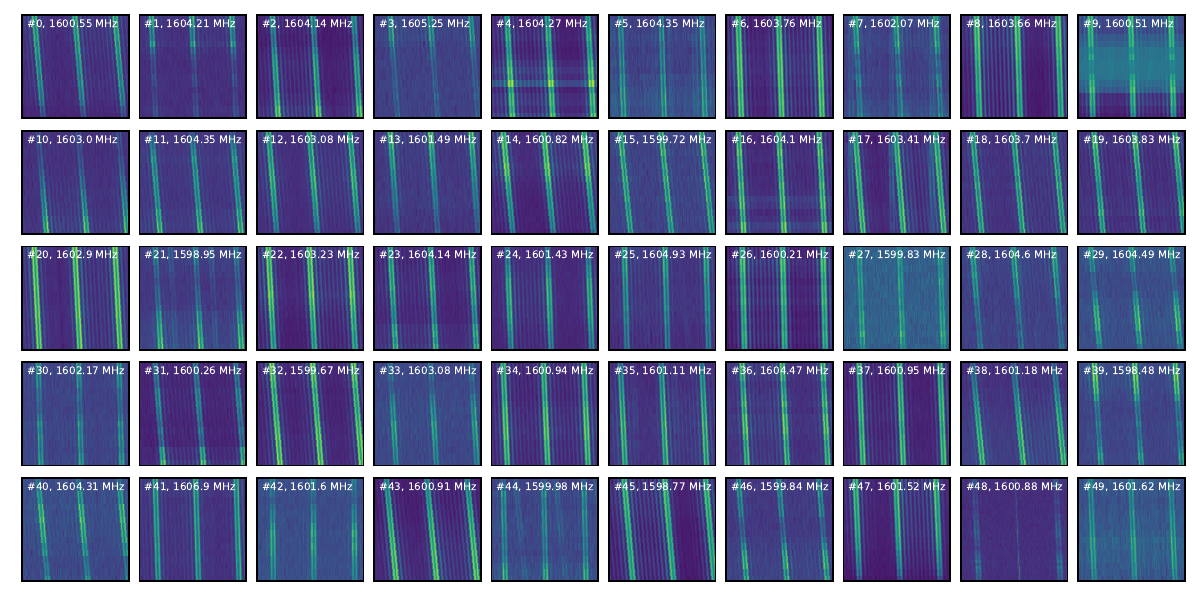}
     \caption{Snippets of dynamic spectra of morphologically-similar signals from GLOBULAR, reproduced from Jacobson-Bell et al.~\cite{jacobsonbell2025}.}
  \label{fig:globular}
\end{figure*}

GLOBULAR also reveals families of related RFI signal morphologies (Fig.~\ref{fig:globular}), and in principle could be used to identify and mitigate sources of interference,
or to correlate with satellite ephemerides to assist with flagging. Outside of Listen, Pinchuk and Margot used supervised ML to create a 
direction-of-origin filter for RFI at GBT \cite{pinchuk2022}.

The vetting that follows a detection can itself be automated. When a narrow-band
signal in observations of Proxima Centauri passed the standard filters in 2019
\cite{smith2021}, it was ruled out only after a lengthy hunt through the
archive, which turned
up dozens of lookalikes at other frequencies, thereby establishing it as RFI
\cite{sheikh2021}. Ma et al.\ built the tool that would have shortened that vetting:
a reverse spectrogram search (akin to a reverse image search), which
encodes a signal of interest with a $\beta$-variational autoencoder and retrieves
the most similar signals from the encoded archive \cite{ma2024reverse}. Given a candidate, the first question is
always whether we have seen it before, and this makes it a question that can be
answered automatically rather than relying on extensive visual inspection.

\subsection{Comparing ON and OFF observations}

Anomaly detection can also compare signals directly between ON and OFF observations, and two Listen studies show how differently
that idea can be realized. Zhang et al.\ started downstream of the
classical search: they ran turboSETI on GBT L-band data and extracted spectrogram snippets around 91,000 of the turboSETI hits
\cite{zhang2019}. On these they trained a generative network, an
adversarial convolutional LSTM, to predict the OFF observation from the
ON observation that preceded it. If the predicted OFF and the real OFF
disagree, something is present in the ON scan that the model could not
account for. 
 
Pardo et al.\ dispensed with the narrow-band detection step
altogether \cite{pardo2025}. They chopped the full band of each
observation into 220-Hz windows, which turns samples of $\sim 800$ stars in
each of three receivers (two from GBT and one from Parkes) into $\sim 10^{11}$ cadence snippets. In the first cut, each
cadence is reduced to fifteen cross-correlation
statistics between its six spectrograms, embedded in two dimensions
with UMAP, and compared against the region of that plane where
setigen-injected signals present only in the ON pointings land. This results in a reduction of the number of snippets by a factor $10^5$.
Two scores then rank what is left, one favoring candidates in relatively RFI-free regions of the spectrum, the
other favoring candidates whose ON observations resemble one another
and not the OFFs. About 20,000 cadences were
visually inspected, including the highest-ranked candidates, along with some validation samples. Although all candidates were rejected as possible technosignatures, the seven orders of magnitude reduction
from the original set of cadence snippets enabled a thorough search 
with no assumption about the signal morphology beyond it fitting within the 220-Hz window.

Anomaly detection does not guarantee that the signals that survive are good technosignature candidates.
Most are pernicious kinds of RFI; just a weirder-looking stone
stuck in your shoe (Fig.~\ref{fig:shoe}). But these kinds of filtering, so long as they also preserve the majority of injected ``interesting'' signals,  
make the manual inspection stage manageable.

This is not a SETI-specific problem. Baron and Poznanski identified the
weirdest galaxy spectra in the Sloan Digital Sky Survey (SDSS) with an unsupervised outlier method
and found a wide variety of weirdness, some of it mundane
\cite{baron2017}. Astronomaly \cite{lochner2021} and the active anomaly
detection framework of Ishida et al.\ \cite{ishida2021} address this by
putting a scientist in the loop: the algorithm proposes, the human
labels, and the ranking learns what this particular investigation
considers interesting rather than merely unusual. In future, LLMs may start to play at least part of the role of the human vetter here (Section~\ref{sec:llm}).

\subsection{Preliminary Approaches to Unfamiliar Structure}

Listen is now exploring teacher--student networks as
a way of scoring unfamiliarity directly (V.~Gajjar, private
communication, 2026). Two networks trained to agree on
familiar inputs will tend to disagree on unfamiliar ones, so their
disagreement is an anomaly score. What constitutes an anomaly does not have to be defined
at the training stage, so the hope is that this approach can find novel signals that are true outliers from the background RFI population.

Rogers et al.\ have taken a different tack with a method they call
\emph{What We Don't C} \cite{rogers2025}. It uses latent flow matching
to strip out the variation in a learned representation that is
explained by a set of supplied conditioning variables, leaving whatever
structure those variables do not account for. For discovery, that is
an appealing capability: it is a way of asking a dataset what it
contains beyond the attributes you already know about. This method is being used to disentangle selection 
biases from astrophysical effects in galaxy spectra \cite{rogersinprep} and also shows promise for searches for
technosignatures and other anomalies.

\section{Real-Time AI at the Telescope\label{sec:realtime}}

Many modern radio telescopes, particularly large-N arrays, generate too much data to permit long-term storage of
the raw voltages.
Something has to decide, close to the moment of acquisition, which subsets of data are of sufficient interest
to save for offline analysis. This makes the
computing architecture part of the search strategy, because once the choice is made, the remaining data are lost.

Within Listen, a graphics-processing-unit (GPU) based streaming framework, Stelline \cite{stelline}, is being developed for the Allen Telescope Array (ATA). It
combines signal processing, visualization, and inference in one
pipeline, so that a specialist neural network can run on the data
being output from the channelization and beamforming (powered by BLADE, the Breakthrough Listen Accelerated Digital Signal Processor (DSP) Engine)
and flag events for further analysis as they happen. 

As a demonstration, BLADE was used to feed data into FRBNN, a system designed to find FRBs in real time at the ATA \cite{ma2025}.
A compact network was trained on simulated bursts injected into real telescope
backgrounds. When run on streaming data, it successfully found dispersed pulses.

The architecture this implies is layered by cost. Cheap models look at
everything; expensive models look at what the cheap ones select. A model that takes seconds to think
about one image cannot look at a telescope data stream, but it can look
at the handful of candidates per hour that a fast detector hands it.

\begin{figure*}[t]
  \centering
  \includegraphics[width=\textwidth]{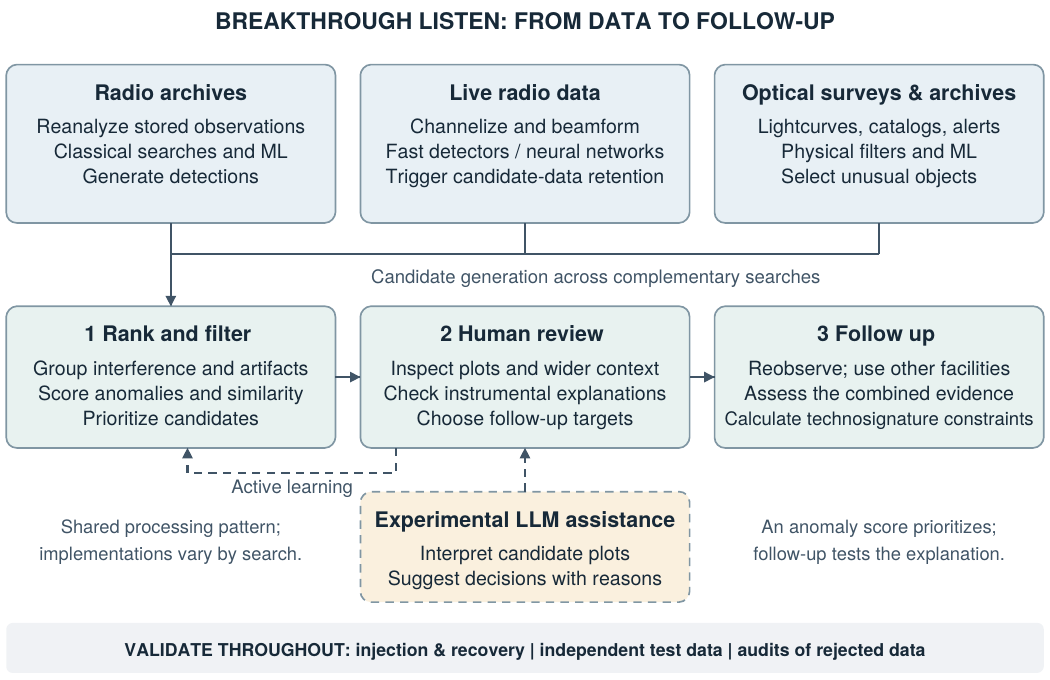}
  \caption{Candidate processing across Breakthrough Listen searches.
  Radio archives, live radio streams, and optical survey products provide
  complementary routes to candidate generation. Classical methods and
  machine learning support filtering and prioritization, followed by
  review and targeted follow-up. Dashed elements denote work in development: multimodal LLM assistance with candidate vetting, and active learning, in which human labels refine the ranking.
  This schematic summarizes a shared processing pattern rather than a
  single deployed pipeline. Injection and recovery tests, independent
  validation data, and audits of rejected data are needed to quantify
  what each stage retains and loses.}
  \label{fig:candidate_processing}
\end{figure*}

Each stage of this filtering chain (Fig.~\ref{fig:candidate_processing}) must be evaluated for
what it loses as well as what it retains. Injection and
recovery tests measure sensitivity across signal
strength, drift rate, bandwidth, and intermittency, with
classical and ML methods compared on the same data. Tests on observations excluded
from training are needed to assess performance across
telescopes, epochs, and RFI environments. For unfamiliar
signal classes, completeness remains conditional on the
populations injected: good recovery of simulated signals
does not establish sensitivity to every possible
technosignature. Real-time systems add an irreversible
choice, since rejected voltages are deleted.
Retaining a small audit sample of rejected data, alongside
model versions and selection thresholds, allows
missed signals and changes in performance to be investigated.
One useful performance measure is the reduction
in processing and human effort at a quantified recovery
rate, rather than candidate reduction alone.

\section{Preliminary Multimodal LLM Work\label{sec:llm}}

Stoppa et al.\ used LLMs to classify optical
transient images and produce textual descriptions of each candidate
\cite{stoppa2025}. Recently, the Listen team has begun to experiment with
multimodal LLMs
for vetting radio technosignature candidates (V.~Gajjar \& L.~Cruz,
private communication). Their SETI Bench framework
compares open- and closed-weight vision--language models
on candidate waterfall plots, asking each model to classify
the signal and explain its decision using criteria such as
its presence in ON versus OFF observations. Preliminary
examples show models correctly rejecting interference
visible across both sets of pointings and articulating the
reason for rejection. Approaches being explored to improve
performance include adding expert guidance to prompts,
providing labeled examples within the prompt (few-shot
learning), and adapting open-weight models through low-rank
adaptation (LoRA). These experiments suggest a role for
multimodal models in reducing the manual vetting burden. The team is testing precision and recall on injected signals to compare performance with the classical approach ahead of widespread deployment.

A combination of approaches may be useful. For example, 
a fast specialist
network like FRBNN can look at the full data stream. A more expensive
multimodal model can interpret its output, perhaps after an
anomaly-detection stage has selected unusual outliers. 
An orchestrating model, or ``agentic observatory'', can turn a
scientific request into a sequence of observing and analysis steps. These systems are all being prototyped at the ATA \cite{stelline}.

\section{Optical Searches\label{sec:optical_ml}}

The next generation of optical survey telescopes, in particular the Vera C.\ Rubin Observatory and its Legacy Survey of Space and Time (LSST), 
is starting to generate huge volumes of data. This includes images and lightcurves for enormous numbers and varieties of objects. 
As in our radio searches, triage is required to find objects of interest (including possible technosignatures).

\subsection{Anomaly Detection in Lightcurves\label{sec:lightcurves}}

Boyajian's Star (Section~\ref{sec:optical}) raises the question of how many anomalies might be hiding in large datasets and are missed by manual inspection.
To automate the search, Giles and Walkowicz used unsupervised learning, reducing each Kepler lightcurve to a set of features and assigning a ``weirdness'' score from a density-based clustering algorithm, so that objects which do not fit into any cluster rise to the top of the list \cite{giles2019}. The method is signal-agnostic by design. It makes no assumption about what a technosignature looks like, only that it will not look like the bulk of the population.


The Listen team has applied an outlier detector built on the Giles and Walkowicz approach to TESS lightcurves from the primary mission, followed by multiple rounds of manual and statistical vetting to remove eclipsing binaries, systematics, and other contaminants, and by injection of simulated megastructure transits to establish detectability thresholds as a function of stellar magnitude, transit duration, and depth \cite{giles2024}. The same weirdness score was used to check the SETI Ellipsoid targets in TESS \cite{cabrales2024}. As with the radio anomaly searches, the output is a ranked catalog of unusual objects, and the vetting stage is still important.

Machine learning also offers a way to infer shapes under an explicit training prior. Mathews et al.\ trained convolutional networks on $10^5$ simulated transits of arbitrary shapes to predict the outline of the occulter directly from the lightcurve, with the aim of flagging Kepler transits whose predicted shape departs substantially from a disk \cite{matthews2024}. 

\subsection{Alert Streams\label{sec:alerts}}

Gallay, Davenport, and Croft looked at how technosignature searches could use real-time alert brokers that ingest data from the
Zwicky Transient Facility (ZTF), and built a workflow for picking out novel high-amplitude
stellar dippers \cite{gallay2025}. The brokers already carry most of what such a search needs, and the filtering can 
be expanded to find other classes of interesting objects.

Iskandarli et al.'s Anomaly Hunter for Alerts (AHA) runs unsupervised
learning on the ZTF alert stream through the Lasair broker
\cite{iskandarli2026}. Three autoencoders work independently, on the
object's tabulated features, on its image cutouts, and on its lightcurve, each trained on a reference population that is mostly ordinary
supernovae. Whatever a given autoencoder reconstructs badly is, by
that autoencoder's standards, unusual. The three autoencoders
did not always agree with one another. An object that looked strange in its light
curve often looked ordinary in its cutouts, and vice versa. Images,
lightcurves, and derived features encode different things, and a
search that looks at only one of them will miss anomalies that are
obvious in another. Running live for 25 days, AHA produced 87 unusual supernova candidates
for follow-up \cite{iskandarli2026}. This method could also in principle be used to 
find technosignature candidates, and human review or agentic AI could be used to filter the output.

\subsection{Solar System Objects and Interstellar Visitors\label{sec:sso}}

Rogers et al.\ applied an autoencoder and a Gaussian mixture model to simulated LSST Solar System object
catalogs,
both to flag outliers in orbital parameters and / or color, and to use the learned representation to
find more objects like a given odd one \cite{rogers2024}. They were able to recover several distinct classes of
objects including the population of simulated interstellar objects (ISOs) in the catalog. 

Interstellar probes have been discussed as a technosignature class for forty years
\cite{freitas1985}, but constraints on their abundance from current surveys are very poor. Rubin is expected to find several ISOs each year, and although an individual ISO is unlikely to be an artificial probe, every one of them
deserves thorough study whatever one's prior \cite{davenport2025}.

3I/ATLAS, the third interstellar object to be discovered, received extensive follow-up from Listen and collaborators.
The Allen Telescope Array observed it within days of
discovery, across 1--9~GHz \cite{sheikh2025}.
MeerKAT observed it for
cometary chemistry, and found hydroxyl from the coma as expected, while
BLUSE searched the same data for narrow-band emission in parallel
\cite{atel17499}. Parkes observed on three epochs with the
Ultra-Wideband Low receiver. Less than 24 hours
before closest approach we pointed the GBT at it with four receivers
spanning 1--12\,GHz, which at that distance is sensitive to a
transmitter of about 0.1\,W equivalent isotropic radiated power, or
roughly a mobile-phone handset \cite{gbt3iatlas}. Listen also took
part in the Rubin observations of the object \cite{chandler2025}.

Nothing in these data suggests that 3I/ATLAS is anything other than a natural object,
but we do know that artificial objects on heliocentric orbits are recoverable by existing
survey pipelines. In January 2025 the Minor Planet Center designated a new near-Earth
object, 2018~CN41, from a linkage of survey tracklets 
in its archive \cite{mpec2025a38}. Shortly thereafter, the orbit was recognized as that of the
Falcon Heavy upper stage and its Tesla Roadster payload, and the designation was
withdrawn \cite{mpec2025a49}. Like Voyager~1 for the radio searches, however, the Roadster is a
bona fide technosignature that current optical pipelines detect, characterize, and
report using nothing more than the tools already in routine use. 

Whether an anomalous object is recovered at all depends on the linking stage, which assumes
Keplerian motion. White et al.\ tested this directly by adding a nongravitational acceleration
to the Rubin survey simulator and running the results through the same HelioLinC3D linking
code the Rubin Solar System pipeline uses \cite{white2026}. Linking succeeds at 96\% or better
of the gravity-only rate for accelerations up to those of the most strongly outgassing known
comets, and only fails when the acceleration is an order of magnitude or more beyond anything
in the natural population. That is good news for the census, and it defines a search: since
at least 98\% of known Solar System objects have nongravitational accelerations below
$10^{-6}$~au\,day$^{-2}$, anything above that threshold that Rubin successfully links deserves
a second look. Anything that \emph{should} have been linked at a given acceleration but was
not is interesting for the opposite reason; an object that maneuvers is precisely the kind
that the linking algorithm is designed to reject, so ``disappearing'' objects are worth
tracking as a candidate class in their own right.

Put together, these studies suggest a workflow: Rubin data products are
filtered for anomalies, the survivors are evaluated against whatever
context is available, and a prioritized list is created for follow-up.
Different branches can use physically motivated cuts, learned anomaly
scores, or both. Follow-up then tests specific hypotheses, including possible technosignatures.

The ZTF and Solar System work supplies the methods and the 3I/ATLAS
campaign supplies an end-to-end test of the final stages. What remains
is the transfer to Rubin at scale, which is not trivial, and the training examples for a survey that has only just
started are, by definition, few. Listen's role in all this is to connect a much larger discovery space to observations and pipelines that
can actually test whether something is technological.

\section{Conclusion}

AI is already doing useful work in Listen: recovering signals that
earlier searches missed, making searches for new signal classes
affordable, cutting the interference haystack by large factors, and
running in real time at the telescope. Anomaly detection in optical
and Solar System data widens the net well beyond traditional radio searches,
and the early LLM experiments
suggest a path toward interpretation and coordination at scale. The
prize is an experiment that can look at more data and more of
the possible signal space than any previous search. 

As our tools and techniques continue to develop, we will continue to measure our progress the way we have always done: 
by the region of parameter space excluded, by reobservation of signals of interest,
and by the weight of evidence we can bring to bear on each candidate
when one finally does not go away. In the new era of artificial intelligence aiding the search for extraterrestrial intelligence, our chances of success have never been better.

\iacunnumbered{Acknowledgements}
The Breakthrough Prize Foundation funds the Breakthrough Initiatives, which manages Breakthrough Listen.

In preparation of this paper, LLMs (Claude and ChatGPT) were used to assist with brainstorming the paper's contents, constructing an outline, and generating Fig.~\ref{fig:candidate_processing}.

\iacunnumbered{References}
\vspace{-14pt}
\renewcommand{\refname}{}
\bibliographystyle{elsarticle-num-ads}
{\fontsize{9}{10.5}\selectfont
\setlength{\bibsep}{0pt}
\bibliography{refs}}

\end{document}